\documentclass[useAMS,usegraphicx,usenatbib]{mn2e}

\usepackage{times}
\usepackage{amssymb}
\usepackage{amsmath}
\usepackage{rotate}
\newif\ifAMStwofonts
\AMStwofontstrue

\def\cloudy{\textsc{cloudy}}
\def\photoion{\textsc{photoion}}
\def\xstar{\textsc{xstar}}
\def\nist{NIST}
\def\xmm{{\it XMM-Newton}}
\def\chandra{{\it Chandra}}

\def\mcg{{MCG--6-30-15}}
\def\mrk766{{Mrk 766}}

\def\cm{{\rm\thinspace cm}}
\def\cts{{\rm\thinspace cts}}
\def\ps{{\rm\thinspace s^{-1}}}
\def\km{{\rm\thinspace km}}
\def\s{{\rm\thinspace s}}
\def\pscm{\hbox{$\cm^{-2}\,$}}

\def\kmps{\hbox{$\km\ps\,$}}

\def\angstrom{$\mathrm{\mathring{A}}$}
\def\ctsps{\hbox{$\cts\s^{-1}\,$}}

\title[The soft X-ray absorption lines of \mcg]{The soft X-ray
  absorption lines of the Seyfert 1 galaxy \mcg}

\author[A. K. Turner et al.]  {\parbox[]{6.in} {A. K.
    Turner$^{1}$\thanks{E-mail: akt21@ast.cam.ac.uk}, A. C.
    Fabian$^{1}$, J. C. Lee$^{2}$ and S. Vaughan$^{1,3}$
    \\
    \footnotesize
    $^{1}$Institute of Astronomy, Madingley Road, Cambridge CB3 0HA\\
    $^{2}$Chandra Fellow, Massachusetts Institute of Technology,
    Center for Space Research, 77 Massachusetts Ave. NE80, Cambridge,
    MA 02139, USA\\
    $^{3}$X-Ray and Observational Astronomy Group, Department of
    Physics and Astronomy, University of Leicester, Leicester, LE1 7RH\\
  }}

\begin{document}

\maketitle

\label{firstpage}

\begin{abstract}
  The absorption lines in the soft X-ray spectrum of \mcg\ are studied
  using the Reflection Grating Spectrometer data from the 2001 \xmm\ 
  320~ks observation. A line search of the full time-averaged spectrum
  reveals 51 absorption lines and one emission line. The equivalent
  widths of the lines are measured and the majority of the lines
  identified. We find lines produced by a broad range of charge states
  for several elements, including almost all the charge states of
  oxygen and iron, suggesting a broad range of ionization parameters
  is present in the warm absorber. The equivalent widths of the lines
  are broadly consistent with the best fitting warm absorber models
  from \citet{turner03}. The equivalent widths of the absorption lines
  allow confidence limits on the column density of the species to be
  determined. For O\thinspace\textsc{vii} a column density of
  $10^{18.36}$--$10^{18.86}$\pscm is found. This column density of
  O\thinspace\textsc{vii}, when combined with the inferred
  Fe\thinspace\textsc{i} absorption, is sufficient to explain the drop
  in flux at 0.7~keV as being due to absorption from the warm
  absorber. Fitting O\thinspace\textsc{i} K-edge absorption to the
  spectrum reveals a column of $10^{17.51}-10^{17.67}$\pscm of
  O\thinspace\textsc{i}, suggesting an Fe:O ratio of $\sim$1:2,
  consistent with the neutral iron being in the form of iron oxide
  dust. Variability is seen in a few absorption lines, but the
  majority of the absorption features, including the prominent
  absorption edges, stay constant throughout the observation despite
  variability in the continuum flux.

\end{abstract}

\begin{keywords}
galaxies: active -- galaxies: Seyfert: general -- galaxies:
individual: \mcg\ -- X-ray: galaxies 
\end{keywords}

\section{Introduction}

\mcg\ is a bright, nearby Seyfert 1 galaxy that shows many of the
typical X-ray spectral features found in other Seyfert 1s. The X-ray
spectrum consists of a power-law continuum with $\Gamma\sim1.9$, broad
iron K$\alpha$ emission line and a soft excess (e.g.
\citealt{fabian02}). The spectrum of \mcg\ also shows a sharp drop in
flux at $\sim$0.7~keV. In the pre-\chandra/\xmm\ era, this drop was
interpreted as being due to O\thinspace\textsc{vii} and
O\thinspace\textsc{viii} K-shell edges from ionized gas in the
environment of the AGN \citep[``the warm
absorber'';][]{otani96,reynolds97sample,george98}.

This interpretation of the drop in flux at 0.7~keV was challenged by
\citet{brand-ray01}, using a $\sim$120~ks Reflection Grating
Spectrometer (RGS) \xmm\ observation. They explain the drop as the
blue wing of a broad O\thinspace\textsc{viii} emission line. The
broadening mechanism of the line was taken to be the same relativistic
blurring that produces a broad iron line at $\sim$6~keV.  Part of the
reason for adopting such a model was an apparent $\sim$16,000~\kmps
discrepancy in the position of the drop in flux and the energy of the
O\thinspace\textsc{vii} edge (0.74~keV). Smaller drops at $\sim$0.39
and $\sim$0.54~keV were similarly explained as being due to emission
from C\thinspace\textsc{vi} and N\thinspace\textsc{vii} respectively.

\citet{lee01} countered these claims with yet higher spectral
resolution High Energy Transmission Grating Spectrometer (HETGS)
observations taken with \chandra. This discrepancy was interpreted by
\citet{lee01} to be due to the presence of an Fe\thinspace\textsc{i}
L-shell absorption edge at the 0.7~keV drop, which when combined with
O\thinspace\textsc{vii} resonance absorption lines and edge could
adequately explain the drop in flux at 0.7~keV. \citet{lee01}
concluded the Fe\thinspace\textsc{i} could be in the form of dust
embedded in the warm absorber (the Dusty Warm Absorber, DWA), either
in the form of iron oxides or silicates, which would explain the
optical observations of \citet{reynolds97} where $E(V-B)$ reddening
suggests strong dust extinction is present (see \citet{ballantyne03}
for a discussion of the location of the dust).

\citet{sako03}, however, fitted the DWA model of \citet{lee01} and the
Relativistic Emission Line (REL) model of \citet{brand-ray01} to the
\xmm\ RGS observation and claimed that the REL model provides a much
better fit to the data than the DWA model. Models of absorption by
various species were also fitted to the absorption lines seen in the
data. \citet{sako03} claimed that the column densities of
O\thinspace\textsc{vii} and O\thinspace\textsc{viii}, derived from the
observed absorption lines, are insufficient to explain the 0.7~keV
drop by absorption.

\mcg\ was observed for a further three \xmm\ revolutions in 2001,
providing 320~ks of data when the source was at a higher flux level
\citep{fabian02} than the previous \xmm\ observation reported by
\citet{brand-ray01} and \citet{sako03}. An analysis of the RGS
spectrum unambiguously showed the presence of Fe\thinspace\textsc{i}
through the absorption fine structure at the L-shell edge
\citep{turner03}. The 0.7~keV drop could also be adequately fitted by
Fe\thinspace\textsc{i} and O\thinspace\textsc{vii} absorption.
Self-consistent models of warm absorption were made using the
photoionization code \cloudy\ and fitted to the data along with the
REL model. Both models provided similar quality fits to the data and
both were capable of explaining the gross features of the spectrum.
The DWA model was, however, more physically self-consistent, since the
REL model requires the underlying continuum to be very flat below
$\sim1$~keV and steep above this energy. Conventional models of
Seyfert 1 emission do not contain such a break.  Moreover, the
predicted equivalent widths of the emission lines of the REL model in
\citet{brand-ray01}, \citet{sako03} and \citet{turner03} were much
larger than the largest equivalent widths predicted by ionized disc
models and also lacked the iron L emission predicted by those same
models \citep[e.g.][]{ballantyne02,rozanska02}.

\mcg\ also shows interesting spectral variability: The flux in the
iron K$\alpha$ line is observed to stay constant while the flux in the
continuum varies \citep*{lee00,vaughan01,lee02,shih02,fabian02}. If a
spectrum from a low flux period is subtracted from a spectrum from a
high flux period then the resulting spectrum is adequately fitted by a
power-law \citep{fabian02}. These properties led \citet{fabian03} to
propose a two component model for the emission in \mcg\, consisting of
a varying power-law component and an almost constant reflection
dominated component. This differencing method was used by
\citet{turner03} to remove the effects of absorption from the 320~ks
RGS observation and determine the underlying continuum. It shows that
the spectrum does not contain any strong relativistic emission lines
and confirms that absorption is responsible for the 0.7~keV drop.

In this paper we use the absorption lines seen in the RGS spectrum
from the 320~ks \xmm\ observation to measure the parameters of
individual species that make up the warm absorber. We do not assume a
continuum, either DWA or REL, but instead use a cubic spline fit to
determine the flux level around the absorption lines.  This way the
absorption line parameters can be determined without any prior
assumption being made about which model is correct. This information
is used to confirm that a DWA model accounts for the observed spectral
features, specifically the drop in flux at 0.7~keV.  The paper is
organised as follows: Section \ref{absorption_lines} describes the
method used to find, identify and measure the absorption lines as well
as the results obtained by this method. In section \ref{dust} the
evidence for the presence and composition of dust is examined. In
section \ref{variability} the variability of the absorption lines is
measured. In section \ref{discussion} the results are discussed.

\section{Absorption line properties}

\label{absorption_lines}

\subsection{Observations and data reduction}

\label{data_reduction}

We observed \mcg\ with \xmm\ on 2001 July 31 to 2001 August 5
(revolutions 301, 302 and 303) for $\sim320$~ks \citep{fabian02}. The
Observation Data Files (ODFs) were analysed using the standard
reduction chains provided by the 5.4.1 release of the \xmm\ Science
Analysis Software ({\sc sas v5.4.1}) and the associated calibration files.
The last few ks of each orbit data were removed since these periods
showed strong background flaring. The final RGS spectra were rebinned
so at least 20 counts were present in each spectral bin and were
fitted to trial models in the X-ray spectral fitting package {\sc
  xspec v11.2} \citep{arnaud96}.

The previous \xmm\ observation of \mcg\ was taken on 2000 July 10-11
and consists of $\sim120$~ks of data \citep{brand-ray01}. It was
reduced using the same prescription as described above.

\subsection{Line finding}

\label{line_find}

To determine the presence of any spectral absorption lines, the
spectra were divided up into a number of regions, each region being
analysed separately. The regions were chosen so they did not span any
of the large drops in flux and contained approximately the same number
of spectral bins (the regions are shown as individual panels in
Fig.~\ref{main_table}). The exception to this was the 0.534-0.714~keV
region which spans the Fe\thinspace\textsc{i} edge at 0.702~keV. This
was done to allow the O\thinspace\textsc{vii} $1s^2-1s5p$ line to be
included in the same region as the other O\thinspace\textsc{vii}
$1s^2-1snp$ lines. In this case a model of Fe\thinspace\textsc{i}
absorption was included \citep{turner03}. Within each region the
spectra were fitted with a power-law, Galactic absorption
\citep*[$N_{\rm H}=4.06 \times 10^{20}$~cm$^{-2}$;][]{elvis89} and an
additional multiplicative model component to account for any unknown
variations in the underlying continuum either from complexity in the
underlying continuum or weak photoelectric edges. A cubic spline
function with a number of knots evenly separated in energy across the
region was chosen for this additional component \citep{numerical}. The
number of knots varied between 5 and 10 depending on the relative size
of the region to the FWHM of the line spread function. A model
absorption line (with Gaussian opacity profile) was then fitted with
fixed energy to the data at regular intervals across the region.  The
separation between points where the line was fitted was made to be
less than the FWHM of the instrumental response to ensure that no
significant lines were missed. Once the line had been moved across the
region it was fitted with its energy free to vary to the point where
the presence of the line improved the $\chi^2$ of the fit the most.
The significance of the improvement in $\chi^2$ upon adding this line
was assessed using the F-test and a 95 per cent confidence limit. If
the line was found to be significant then the line position was
recorded and the region around this line was removed (within 1
instrumental FWHM either side of the line centre). The process was
repeated until no more points produced significant improvements in the
fit. The process produced a list of 51 lines whose positions span the
whole RGS passband (see Table \ref{main_table}). Because of the
significance level used above, a few of the weaker found lines may be
spurious detections.

\subsection{Line parameter determination}

\label{line_parameter_determination}

Once the positions of the significant lines had been determined, the
equivalent widths of these lines were determined. To determine the
equivalent widths, the regions of the spectra were again individually
fitted to a power-law, Galactic absorption and cubic spline (and
Fe\thinspace\textsc{i} in the case of the 0.534-0.713~keV region).
Then absorption line models were placed at the energies found in
section \ref{line_find} and fitted to the data.

It was decided to use a simple notch model for the line model rather
than the more complicated Gaussian opacity profile model used in
Section \ref{line_find}. A notch model is an absorption model in which
the transmission is zero within the line width and unity outside. The
{\sc xspec} fitting package convolves this absorption model with the
instrumental response before fitting it to the data.  It was found
that this notch model could be adequately fitted to the data for all
but one of the line positions\footnote{This line (line 43) was fitted
  with a model line with a Gaussian opacity profile.  The width of the
  line was determined to be $1870\pm220$~\kmps.}.  This suggests that
the widths of the absorption lines are less than the width of the
instrumental line spread function. The notch model has the advantage
that its width gives a direct measurement of the equivalent width of
the absorption line and so error determination on the equivalent width
is more straight forward. It also provides a good approximate model
for an absorption line lying on the saturated (flat) part of the curve
of growth. The accuracy of this model at reproducing absorption lines
that have been convolved with the instrumental line spread function
was tested. The model was found to accurately reproduce the profile
shape for a broad range of line parameters.

The best fits to the regions have reduced $\chi^2$ values of
$\sim1.1$. The best fitting continuum model does not represent the
true emission from the central source as the amount of continuum
absorption from photoelectric edges at any given point in the spectrum
is unknown. Also the details of reflection spectra are highly
uncertain and any attempt to introduce a self-consistent model of the
continuum will result in the continuum level being at the incorrect
level for many of the lines. The purpose of this work is to measure
the absorption line properties without assuming any physical
continuum, either of the DWA or REL type. The continuum model here
represents the flux level to either side of the narrow absorption
lines and allows the equivalent width of those lines to be measured.

The errors on the equivalent widths of the lines were determined using
a standard $\Delta\chi^2=1.0$ criterion. Due to the corrugated nature
of the $\chi^2$ space with respect to line position, a Monte Carlo
technique was used to determine the line position error and avoid
problems with the {\sc xspec} error command. The number of counts
in each spectral bin in the source and background spectra were
perturbed by a Poisson distribution with mean counts equal to the
source and background count rate respectively. The best fitting line
model was then fitted to this new spectra and the resulting best
fitting parameters recorded. This was repeated 1000 times and the
resulting distribution of parameters used to determine the error.

The best fitting line models are shown compared to the combined and
fluxed spectrum from all three orbits in Fig.~\ref{figure_all_1} and
the derived line parameters are given in Table \ref{main_table}. Also
found during the fitting was a narrow emission feature at 0.5565~keV
with an equivalent width of $0.777\pm0.238$~eV and is identified as
O\thinspace\textsc{vii} forbidden emission. The line finding method of
section \ref{line_find} is only able to find the highest equivalent
width lines due to the presence of noise in the data.  Therefore, when
the continuum and line model is fitted to the data, the continuum
component will be below the true level, since it is actually fitting
to the average level of the continuum plus the undetected, low
equivalent width lines.  Since the model continuum is below the true
continuum the equivalent widths of the model lines will be
underpredicted in general. The use of a many knot cubic spline to
model the continuum means that only those missed lines near a found
line affect its equivalent width rather than an average of all the
missed lines in the region. Unresolved blends of many lines will also
be missed and filled in by the spline continuum. This is true for the
iron Unresolved Transition Array \citep[UTA, e.g.][]{sako01,behar01}
which has the effect of lowering the continuum between $\sim0.7$ and
$\sim0.8$~keV.

\begin{figure*}
  \rotatebox{270}{\resizebox{!}{0.97\textwidth}{
      \includegraphics{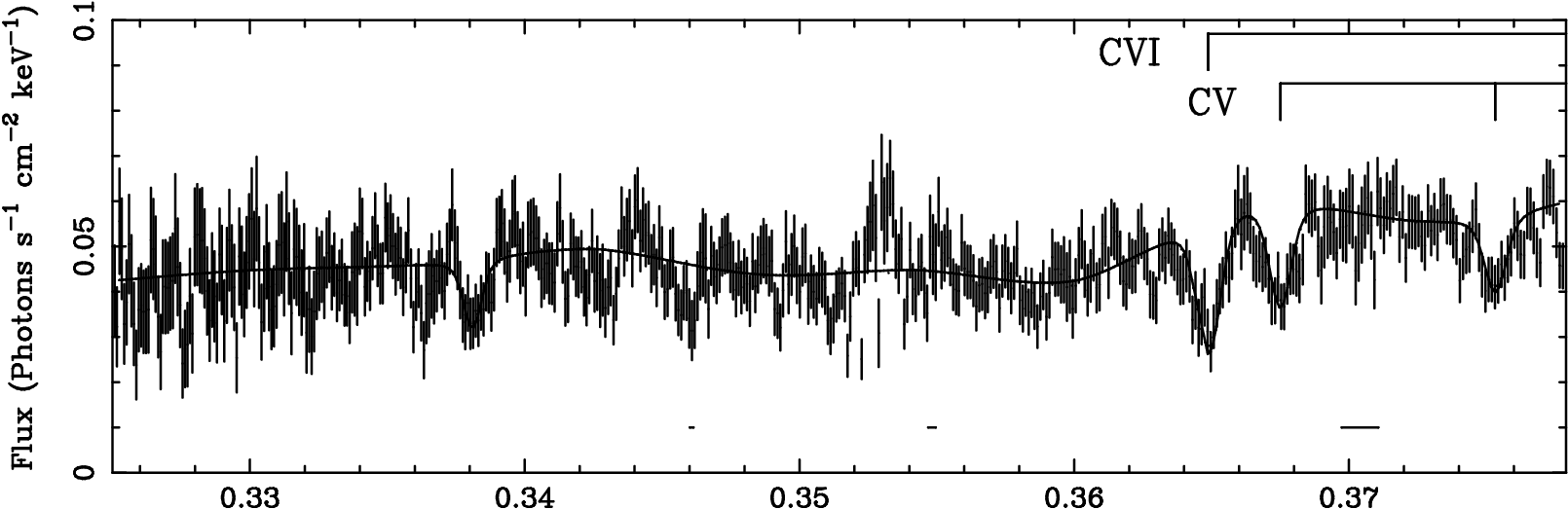}
      \includegraphics{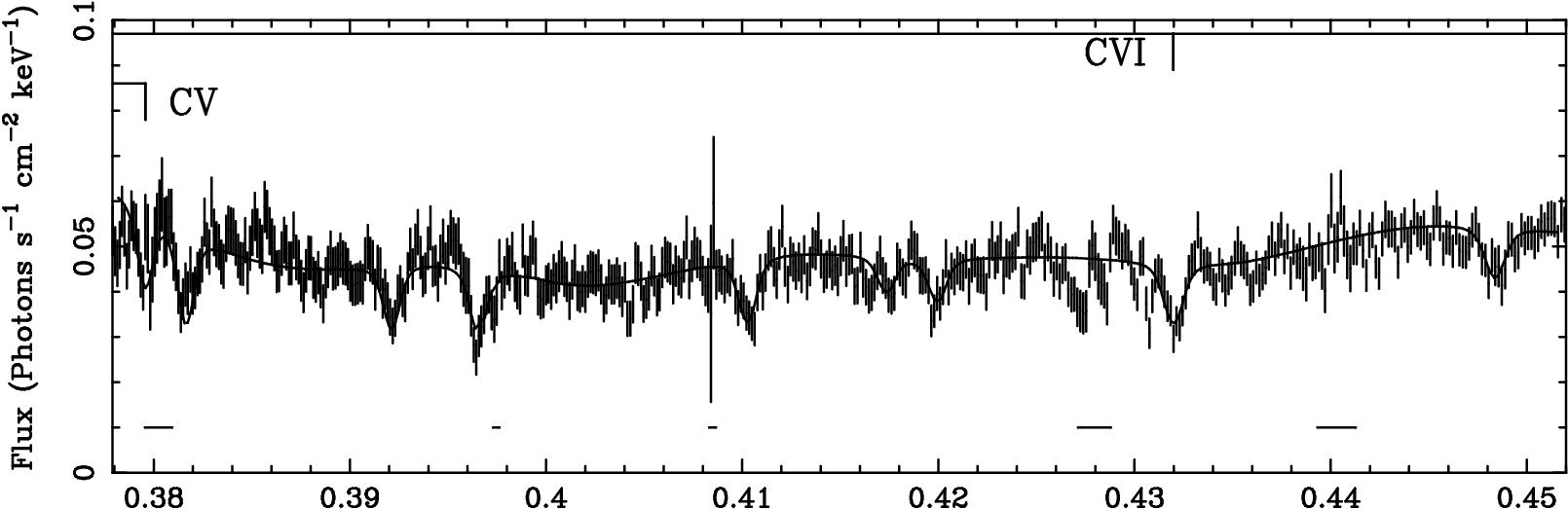}
      \includegraphics{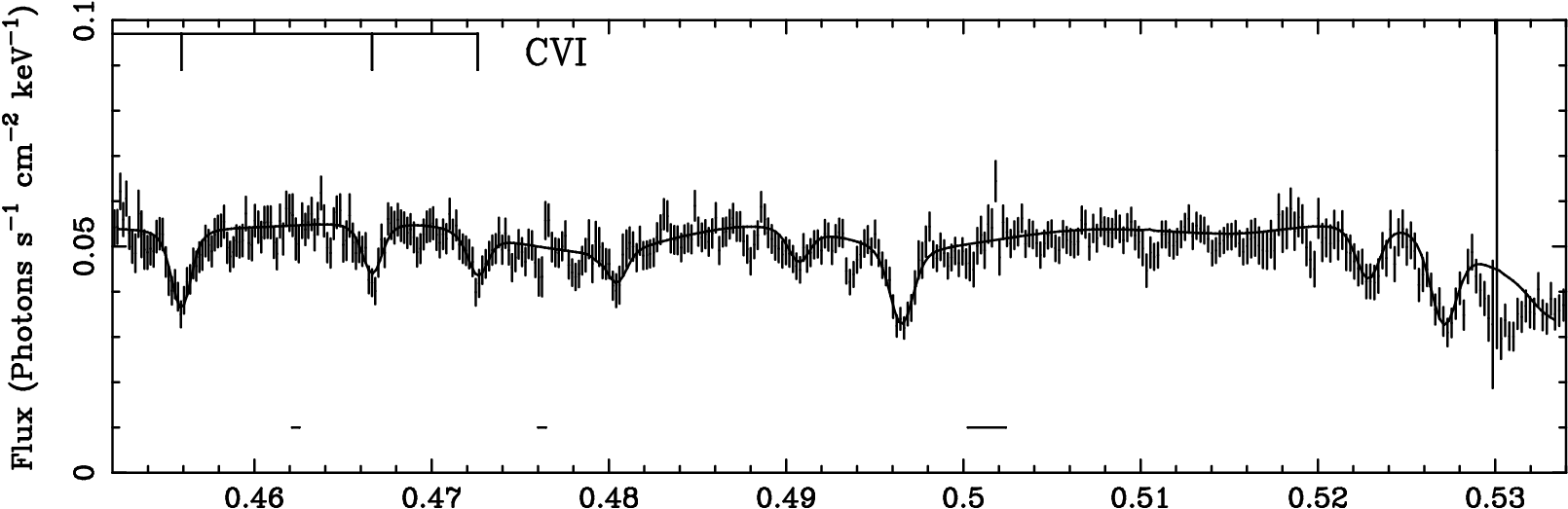}
      \includegraphics{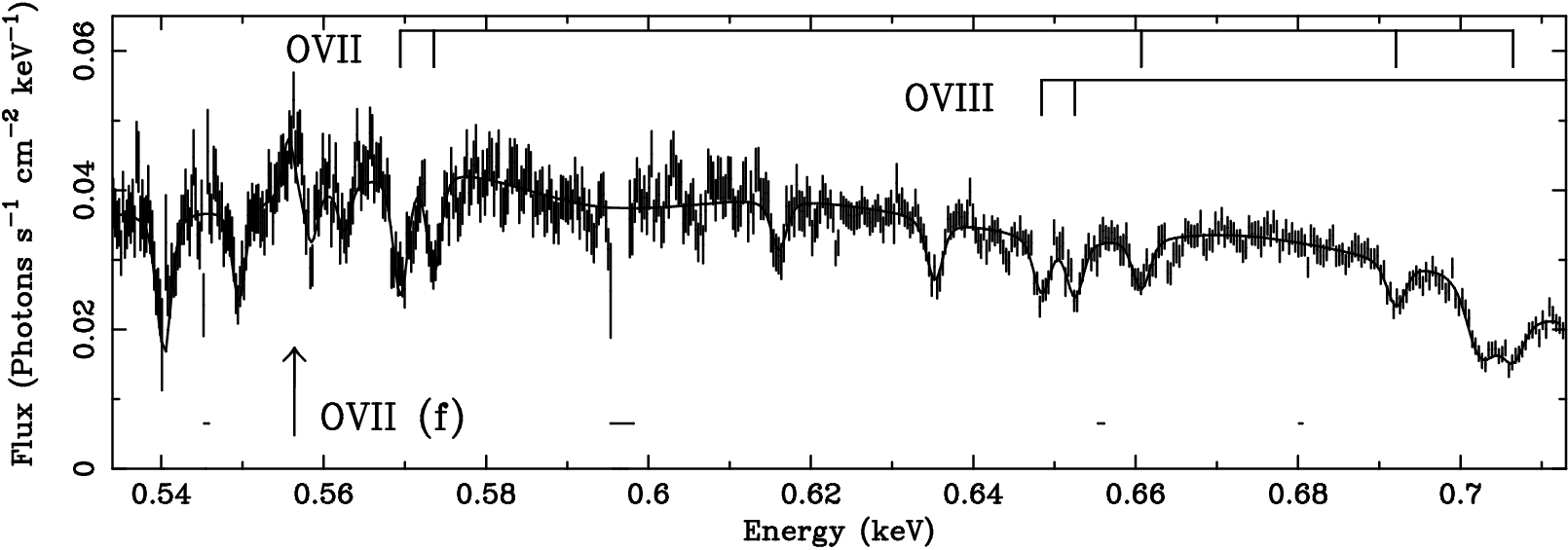}}}
\caption{Best fitting line model compared to fluxed and combined RGS
  spectrum. The lines below the spectrum indicate the positions of
  chip-gaps and other poor quality regions in the detectors. The line
  labels above the spectrum indicate the positions of the lines for
  several species. Also labelled is the O\thinspace\textsc{vii}
  forbidden emission line at 0.5565~keV.}
\label{figure_all_1}
\end{figure*}
\begin{figure*}
\setcounter{figure}{0}
  \rotatebox{270}{\resizebox{!}{0.97\textwidth}{
      \includegraphics{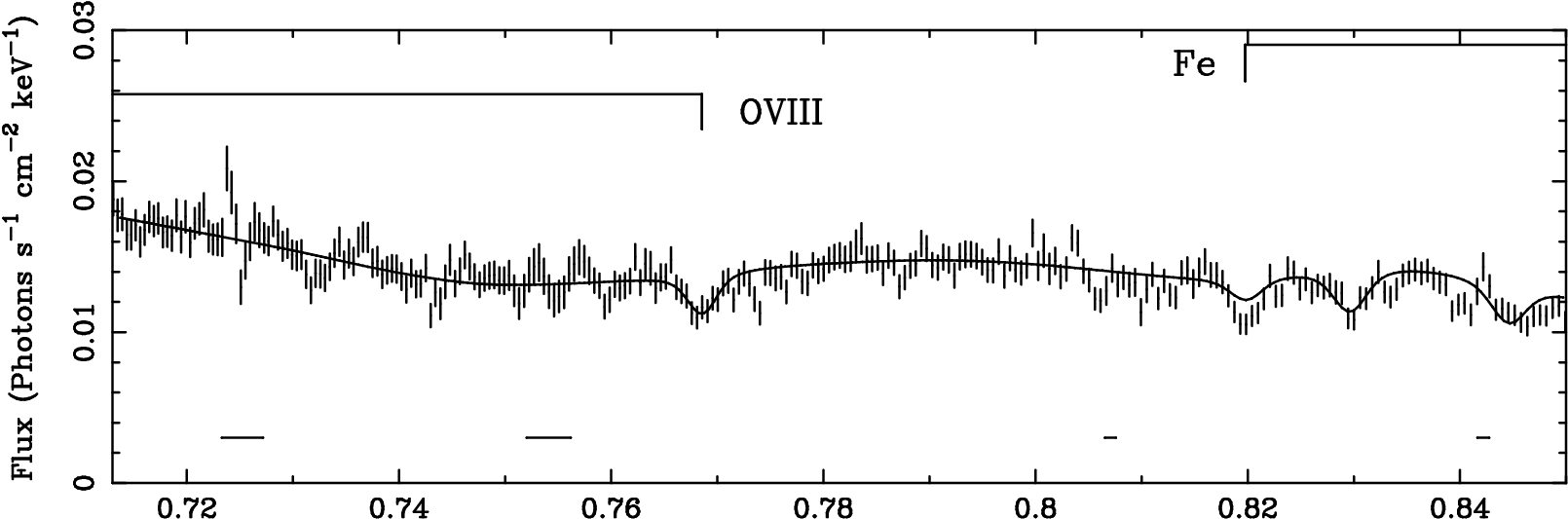}
      \includegraphics{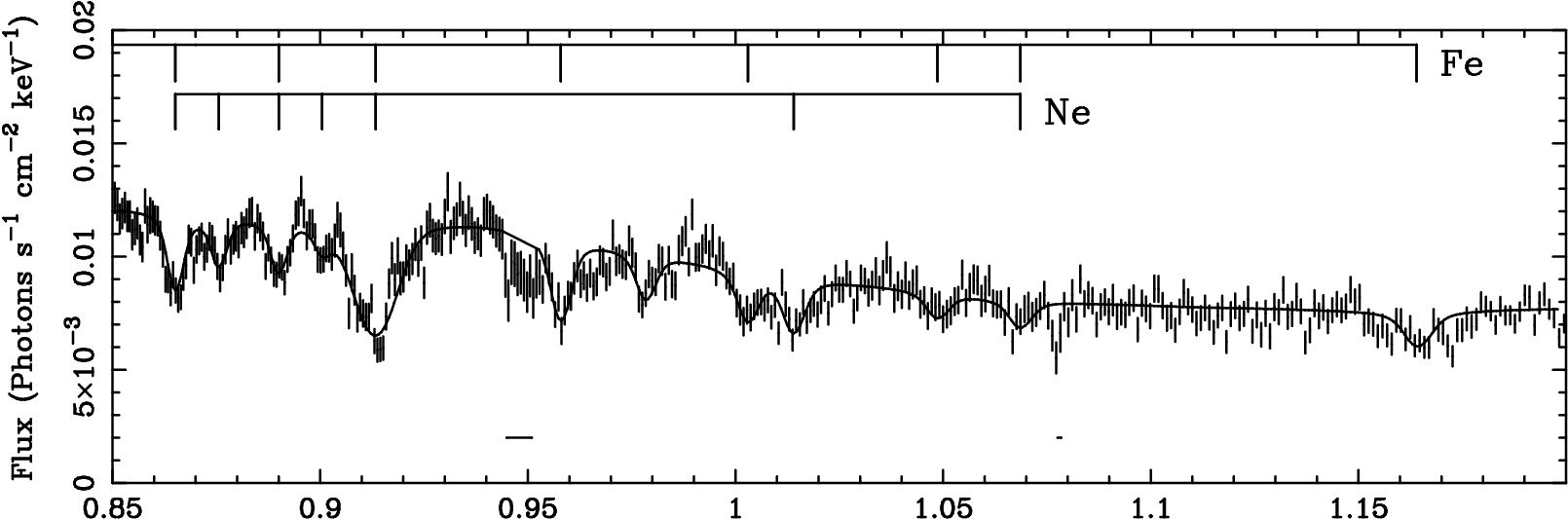}
      \includegraphics{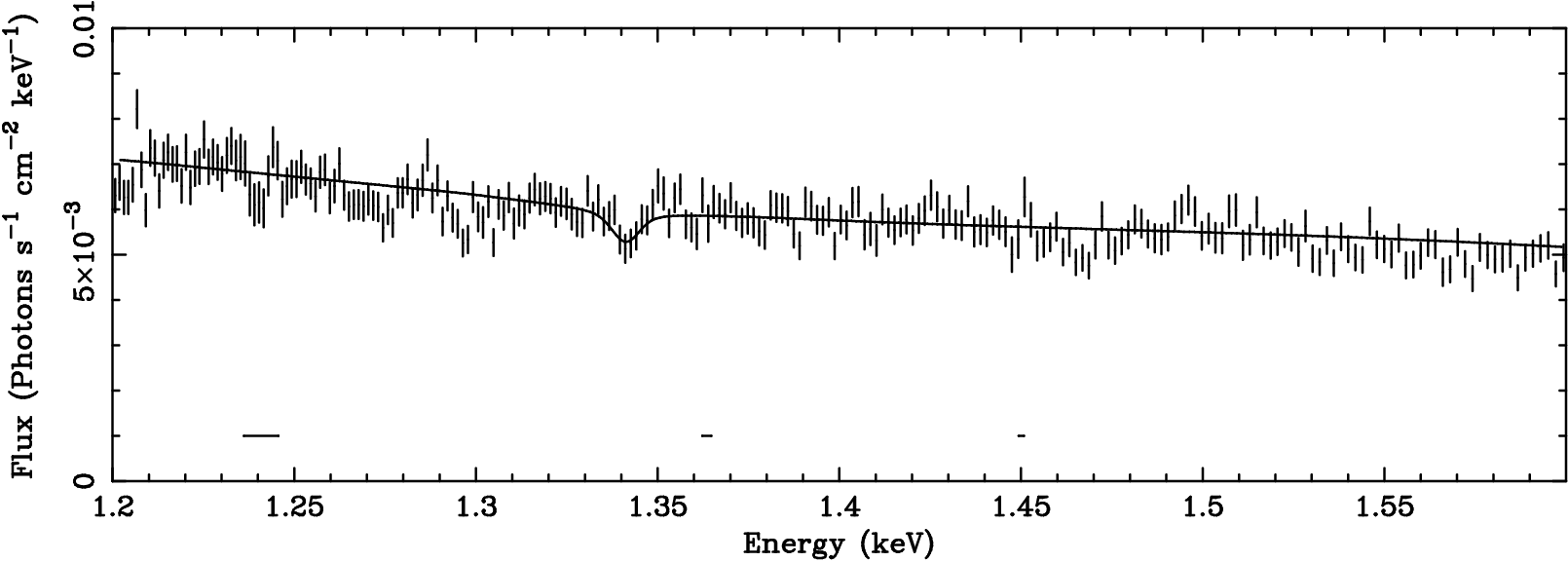}}}
\caption{\emph{Continued}}
\label{figure_all_2}
\end{figure*}

\begin{table*}
\centering
\caption{Table showing lines found in the spectra with measured
  energies, wavelengths, equivalent widths and identifications where they have been
  made. For lines with identifications the rest frame energy and its
  reference are shown (for these identification in brackets if more
  than one is present for a single line), as well as, the velocity
  relative to the frame of \mcg. Reference key: 1: \citet{verner96}, 2:
  \citet{behar02}, 3: \citet{pradhan03}, 4: \citet{mclaughlin98}.}
\begin{tabular}{cccccccc}
\hline
No. & Energy & Wavelength & EW & Identification & Rest Energy & Reference & Velocity \\
 & keV & \AA & eV & & keV & & \kmps \\
\hline
\hline
1 & $0.338090\pm0.000095$  & $36.672\pm0.010$ & $0.372\pm0.070$ &  & & \\
2 & $0.364880\pm0.000076$  & $33.9794\pm0.0071$ & $0.749\pm0.071$ & C\thinspace\textsc{vi} Ly$\alpha$  & 0.367517 & 1 & $-157\pm63$ \\
3 & $0.36751\pm0.00012$    & $33.736\pm0.011$ & $0.498\pm0.059$ & C\thinspace\textsc{v} $1s^2-1s4p$    & 0.370923 & 1 & $461\pm96$ \\
4 & $0.37532\pm0.00011$    & $33.034\pm0.010$ & $0.407\pm0.060$ & C\thinspace\textsc{v} $1s^2-1s5p$    & 0.378533 & 1 & $243\pm92$ \\
5 & $0.37957\pm0.00017$    & $32.664\pm0.014$ & $0.399\pm0.092$ & C\thinspace\textsc{v} $1s^2-1s6p$    & 0.382673 & 1 & $124\pm133$ \\
6 & $0.381635\pm0.000088$  & $32.4876\pm0.0075$ & $0.547\pm0.075$ & S\thinspace\textsc{xiii} $2s^2-2s3p$   & 0.384546 & 1 & $-37\pm70$ \\
7 & $0.39213\pm0.00010$    & $31.6185\pm0.0082$ & $0.433\pm0.071$ & Ar\thinspace\textsc{xii}           & 0.395210 & 1 & $34\pm79$ \\
8 & $0.39654\pm0.00011$    & $31.2668\pm0.0085$ & $0.469\pm0.087$ & Si\thinspace\textsc{xii}           & 0.399752 & 1 & $108\pm82$ \\
9 & $0.41028\pm0.000096$   & $30.2198\pm0.0071$ & $0.461\pm0.072$ &  & & \\
10 & $0.41735\pm0.00046$   & $29.707\pm0.033$ & $0.260\pm0.075$ &  & & \\
11 & $0.41995\pm0.00024$   & $29.524\pm0.017$ & $0.321\pm0.073$ &  & & \\
12 & $0.431980\pm0.000081$ & $28.7014\pm0.0054$ & $0.511\pm0.064$ & C\thinspace\textsc{vi} Ly$\beta$ & 0.435562 & 1 & $164\pm58$ \\
13 & $0.44843\pm0.00013$   & $27.6486\pm0.0081$ & $0.452\pm0.071$ &  & & \\
14 & $0.455880\pm0.000074$ & $27.1967\pm0.0044$ & $0.602\pm0.060$ & C\thinspace\textsc{vi} Ly$\gamma$ & 0.459379 & 1 & $-22\pm49$ \\
15 & $0.466640\pm0.000093$ & $26.5696\pm0.0053$ & $0.408\pm0.058$ & C\thinspace\textsc{vi} Ly$\delta$ & 0.470402 & 1 & $94\pm60$ \\
16 & $0.47260\pm0.00014$   & $26.2345\pm0.0076$ & $0.339\pm0.062$ & C\thinspace\textsc{vi} Ly$\epsilon$ & 0.476391 & 1 & $81\pm88$ \\
17 & $0.48048\pm0.00015$   & $25.8042\pm0.0082$ & $0.316\pm0.078$ &  & & \\
18 & $0.49072\pm0.00028$   & $25.266\pm0.014$ & $0.284\pm0.063$ & Ar\thinspace\textsc{xii}           & 0.495149 & 1 & $382\pm170$ \\
19 & $0.496510\pm0.000064$ & $24.9711\pm0.0032$ & $0.806\pm0.066$ & N\thinspace\textsc{vii} Ly$\alpha$     & 0.500324 & 1 & $-21\pm39$ \\
20 & $0.52284\pm0.00012$   & $23.7136\pm0.0055$ & $0.594\pm0.093$ & O\thinspace\textsc{i}             & 0.526790 & 4 & $-58\pm70$ \\
21 & $0.52704\pm0.00010$   & $23.5246\pm0.0046$ & $0.934\pm0.099$ & O\thinspace\textsc{ii}             & 0.531893 & 3 & $437\pm59$ \\
22 & $0.54035\pm0.00013$   & $22.9452\pm0.0056$ & $1.67\pm0.14$ & O\thinspace\textsc{iv}             & 0.544507 & 3 & $-17\pm74$ \\
23 & $0.54959\pm0.00013$   & $22.5594\pm0.0053$ & $1.123\pm0.097$ & O\thinspace\textsc{v}             & 0.553996 & 3 & $80\pm71$ \\
24 & $0.558430\pm0.00127$  & $22.202\pm0.051$ & $0.652\pm0.086$ & O\thinspace\textsc{vi} KLL            & 0.562287 & 3 & $-253\pm689$ \\
25 & $0.56255\pm0.00017$   & $22.0397\pm0.0065$ & $0.578\pm0.091$ & O\thinspace\textsc{vi} KLL            & 0.567433 & 3 & $279\pm89$ \\
26 & $0.56944\pm0.00013$   & $21.7730\pm0.0051$ & $1.140\pm0.091$ & O\thinspace\textsc{vii} $1s^2-1s2p$    & 0.573956 & 1 & $54\pm71$ \\
27 & $0.57358\pm0.00013$   & $21.6159\pm0.0047$ & $1.006\pm0.089$ & O\thinspace\textsc{vii} $1s^2-1s2p$    & 0.573956 & 1 & $-2127\pm65$ \\
28 & $0.61606\pm0.00016$   & $20.1253\pm0.0052$ & $0.62\pm0.10$ & O\thinspace\textsc{v} & & \\
29 & $0.635260\pm0.000092$ & $19.5171\pm0.0028$ & $0.892\pm0.067$ & O\thinspace\textsc{vi} & & \\
30 & $0.64840\pm0.00011$   & $19.1216\pm0.0031$ & $0.910\pm0.077$ & O\thinspace\textsc{viii} Ly$\alpha$     & 0.653625 & 1 & $92\pm49$ \\
31 & $0.65252\pm0.00012$   & $19.0008\pm0.0036$ & $0.950\pm0.076$ & O\thinspace\textsc{viii} Ly$\alpha$     & 0.653625 & 1 & $-1816\pm57$ \\
32 & $0.66070\pm0.00021$   & $18.7656\pm0.0059$ & $0.876\pm0.086$ & O\thinspace\textsc{vii} $1s^2-1s3p$    & 0.665558 & 1 & $-119\pm95$ \\
33 & $0.69205\pm0.00020$   & $17.9155\pm0.0053$ & $0.942\pm0.087$ & O\thinspace\textsc{vii} $1s^2-1s4p$    & 0.697802 & 1 & $168\pm89$ \\
34 & $0.70651\pm0.00017$   & $17.5488\pm0.0043$ & $1.11\pm0.12$ & O\thinspace\textsc{vii} $1s^2-1s5p$    & 0.712724 & 1 & $313\pm74$ \\
35 & $0.76855\pm0.00035$   & $16.1322\pm0.0073$ & $1.04\pm0.19$ & O\thinspace\textsc{viii} Ly$\beta$     & 0.774623 & 1 & $46\pm138$ \\
36 & $0.81975\pm0.00042$   & $15.1245\pm0.0077$ & $0.71\pm0.22$ & Fe\thinspace\textsc{xvii} & & \\
37 & $0.82959\pm0.00034$   & $14.9452\pm0.0061$ & $1.19\pm0.21$ &  & & \\
38 & $0.84465\pm0.00043$   & $14.6788\pm0.0075$ & $1.418\pm0.021$ &  & & \\
39 & $0.86513\pm0.00037$   & $14.3313\pm0.0061$ & $2.12\pm0.21$ & Fe\thinspace\textsc{xviii}, (Ne\thinspace\textsc{v})      & 0.870740 & 2 & $-379\pm129$ \\
40 & $0.87559\pm0.00053$   & $14.1601\pm0.0086$ & $1.31\pm0.20$ & Ne\thinspace\textsc{vi}            & 0.884340 & 2 & $673\pm183$ \\
41 & $0.89000\pm0.00050$   & $13.9308\pm0.0077$ & $1.54\pm0.20$ & Fe\thinspace\textsc{xix}, (Ne\thinspace\textsc{vii})      & 0.897530 & 2 & $213\pm167$ \\
42 & $0.90043\pm0.00053$   & $13.7694\pm0.0081$ & $0.76\pm0.22$ & Ne\thinspace\textsc{viii}            & 0.908580 & 2 & $390\pm178$ \\
43 & $0.91337\pm0.00043$   & $13.5744\pm0.0065$ & $7.63\pm0.84$ & Fe\thinspace\textsc{xix}, (Ne\thinspace\textsc{ix})      & 0.922023 & 1 & $518\pm144$ \\
44 & $0.95794\pm0.00050$   & $12.9427\pm0.0068$ & $3.46\pm0.32$ & Fe\thinspace\textsc{xx} & & \\
45 & $0.97841\pm0.00063$   & $12.6720\pm0.0082$ & $2.06\pm0.33$ &  & & \\
46 & $1.00300\pm0.00078$   & $12.3613\pm0.0096$ & $2.47\pm0.38$ & Fe\thinspace\textsc{xxi} & & \\
47 & $1.0140\pm0.0011$     & $12.227\pm0.013$ & $2.88\pm0.40$ & Ne\thinspace\textsc{x} Ly$\alpha$   & 1.021810 & 1 & $-14\pm321$ \\
48 & $1.0486\pm0.0012$     & $11.824\pm0.013$ & $1.49\pm0.44$ & Fe\thinspace\textsc{xxii} & & \\
49 & $1.0686\pm0.0017$     & $11.602\pm0.018$ & $1.81\pm0.45$ & Fe\thinspace\textsc{xxiii}, (Ne\thinspace\textsc{ix})      & 1.073783 & 1 & $-869\pm477$ \\
50 & $1.1640\pm0.0032$     & $10.652\pm0.029$ & $3.21\pm0.58$ & Fe\thinspace\textsc{xxiv} & & \\
51 & $1.34100\pm0.00095$   & $9.2457\pm0.0065$ &  $2.31\pm0.51$ & Mg\thinspace\textsc{xi} $1s^2-1s2p$  & 1.352253 & 1 & $192\pm214$ \\
\hline
\end{tabular}
\label{main_table}
\end{table*}

\subsection{Line identification}

%http://physics.nist.gov/cgi-bin/AtData/main_asd
%http://www.nublado.org/
%http://xmm.astro.columbia.edu/photoion/photoion.html
%http://heasarc.gsfc.nasa.gov/docs/software/xstar/xstar.html

The lines were identified by comparing their positions to the energies
quoted in several atomic data sources for the positions of known
absorption transitions. Those sources used were \citet{verner96}, the
\nist\footnote{http://physics.nist.gov/cgi-bin/AtData/main\_asd}
database and the line lists from the
\cloudy\footnote{http://www.nublado.org},
\photoion\footnote{http://xmm.astro.columbia.edu/photoion/photoion.html}
and
\xstar\footnote{http://heasarc.gsfc.nasa.gov/docs/software/xstar/xstar.html}
photoionization codes. For inner-shell transitions
\citet{mclaughlin98}, \citet{behar02}, \citet{behar02conf} and
\citet{pradhan03} were used.  Although \citet{verner96} provides the
most accurate data (it only includes transition energies that have
been experimentally verified) it does not include many important
transitions. One example of this is for the highest oscillator
strength line of Fe\thinspace\textsc{xix} which is not present in the
\citet{verner96} list. The positions for this line in the \nist,
\cloudy, \photoion\ and \xstar\ line lists are 0.9170, 0.9211, 0.9170
and 0.9030~keV respectively. Such variation in energy and the lack of
a complete and accurate line list makes the correct identification of
the observed lines difficult.

In order to identify the observed lines, the transition with the
largest oscillator strength from each of the species considered was
compared to the positions of the observed lines. The species
considered were C\thinspace\textsc{v-vi}, N\thinspace\textsc{v-vii},
O\thinspace\textsc{i-viii}, Ne\thinspace\textsc{ii-ix},
Si\thinspace\textsc{x-xii}, S\thinspace\textsc{x-xiv},
Ar\thinspace\textsc{ix-xvi} and Fe\thinspace\textsc{xvi-xxiv}. Since
there is such a large variation in the theoretical line positions
between the atomic data sets, the transition positions from each
atomic data set were compared simultaneously. If an observed line was
close to the transition from the data sets and it was reasonable to
expect to see it (the transition had a high oscillator strength and
the element has high cosmic abundance or the transition had been seen
in other sources) then the identification was made. Once the highest
oscillator strength line of a particular species had been identified,
the search proceeded for the lines of the same species in order of
decreasing oscillator strength.  If a transition could not be found
then any transitions with a lower oscillator strength from the same
species were not identified.  The exception to this rule was
C\thinspace\textsc{v} where the $1s^2-1snp$ with $n=3,4,5$ lines could
be clearly identified and are expected and the $1s^2-1s2p$ line would
have its location in a confused part of the spectrum at $0.3518$~keV.
No reliable atomic data exists for the inner shell transitions of
nitrogen although they are predicted to occur in the DWA models of
\citet{turner03}. Some of the unidentified transitions around 0.42~keV
may be due to these transitions.

Lines from a broad range of ionization states were identified in the
data. Oxygen and iron states are the most numerous with detections of
almost all species of oxygen and iron species from
Fe\thinspace\textsc{xvii} to Fe\thinspace\textsc{xxiv}, as well as
Fe\thinspace\textsc{i} from the L-edge and Fe\thinspace\textsc{ii} to
Fe\thinspace\textsc{xvi} from the UTA \citep{turner03}. At energies
below $\sim$0.5~keV absorption lines of carbon and nitrogen are
significant. A broad range of neon species are also detected from
Ne\thinspace\textsc{v} to Ne\thinspace\textsc{x} including several
inner-shell transitions. A careful examination of the data shows that
several transitions may have been missed because of chip gaps and
other problems in the instrumental responses. There appears to be a
line directly over a chip gap at 0.427~keV which fits the position of
N\thinspace\textsc{vi}. Another line appears to exist at 0.5306~keV,
whose position matches the position of O\thinspace\textsc{iii}. If
this line is due to O\thinspace\textsc{iii} then every species of
oxygen is evident in the spectrum suggesting a broad range of
ionization parameters is present in the warm absorber. Two velocity
zones are identified, as was found in \citet{sako03}. One has a
velocity of $80\pm260$\kmps and the other $-1970\pm160$\kmps, where
the errors are calculated from the variance of the individual velocity
values for the lines. Only two lines (O\thinspace\textsc{vii}
$1s^2-1s2p$ and O\thinspace\textsc{viii} Ly$\alpha$) are identified
with the second, outflowing, zone. Although not all the observed lines
have identifications some of these may be random Poisson fluctuations
in the data rather than true lines.

\section{warm absorber properties}

\label{warm_params}

\subsection{Confidence region determination}

\label{confidence_region_determination}

In Section \ref{line_parameter_determination} the equivalent widths of
the lines were measured in an independent way with no reference to
which species might be causing those lines or assumptions about the
underlying continuum. Now that those lines have been identified we can
use those equivalent widths, combined with a knowledge of the atomic
parameters of those lines, to calculate confidence regions for the
turbulent velocity of the species and its absorbing column density
towards the central source of X-rays. The profile of an absorption
line which has been Doppler broadened is given by the convolution of a
Lorentzian profile (caused by radiation damping) and a Gaussian
profile (from Doppler broadening) giving the Voigt profile. The
Doppler broadening is caused by a combination of thermal broadening
and turbulent broadening, with turbulent broadening being the dominant
mechanism for the gas within the single zone warm absorber
\citep{nicastro99}. The nature of the Voigt profile means that for a
single line turbulent velocity and column density are highly
degenerate. To break this degeneracy several other lines from the same
species are required or upper limits for the equivalent widths if no
other lines are positively identified. Having more than one line for a
species, therefore, allows tighter constraints to be placed on the
column density and turbulent velocity than would otherwise be the
case. For instance, if the O\thinspace\textsc{vii} $1s^2-1s2p$ line is
considered on its own the column density is only constrained to lie in
the range $\sim10^{16.5}$--$10^{18.8}$\pscm and the turbulent velocity
is unconstrained in the 10--350~\kmps range. The expected equivalent
width of a line given a particular turbulent velocity and column
density can be found by numerically integrating the absorbed area
under the normalised Voigt profile.  Confidence regions for the
turbulent velocity and column density of a particular species were
determined by calculating the expected equivalent widths for all the
observed lines of that species across a grid of turbulent velocity and
column density values.  These values were then compared to the
observed equivalent widths and errors calculated in section
\ref{line_parameter_determination} to produce a grid of $\chi^2$
values. By drawing contours at particular values of $\chi^2$
confidence regions could be calculated. The atomic data used to
calculate the Voigt profiles was from \citet{verner96}, the same
source of atomic data used by \citet{sako03}.

This method can be checked by replacing the notch models with Voigt
profiles for the lines of interest (constraining the column density
and turbulent velocity to be the same for all of the lines) and
fitting to the data.  Confidence regions are then obtained by
calculating the $\Delta\chi^2$ from the best fit for a grid of column
density and turbulent velocity.

\subsection{Results}

\subsubsection{Oxygen}

\begin{figure}
\rotatebox{270}{
\resizebox{!}{\columnwidth}
{\includegraphics{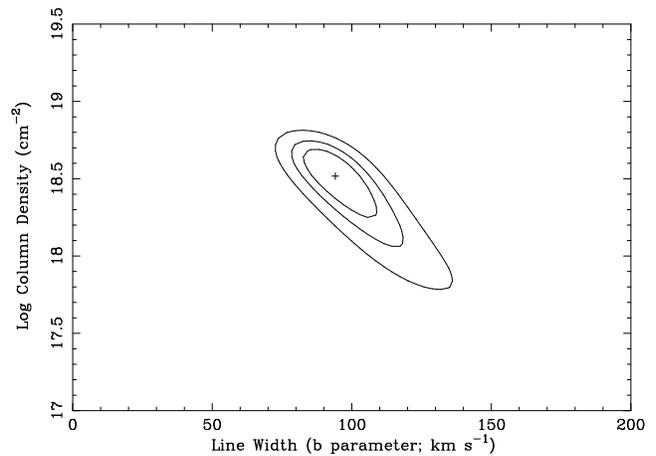}}}
\caption{Confidence region for column density and turbulent velocity
  for O\thinspace\textsc{vii} using measured equivalent widths of the 
  O\thinspace\textsc{vii} $1s^2-1snp$, $n=2,3,4,5$ absorption lines.}
\label{ovii_1}
\end{figure}

\begin{figure}
\rotatebox{270}{
\resizebox{!}{\columnwidth}
{\includegraphics{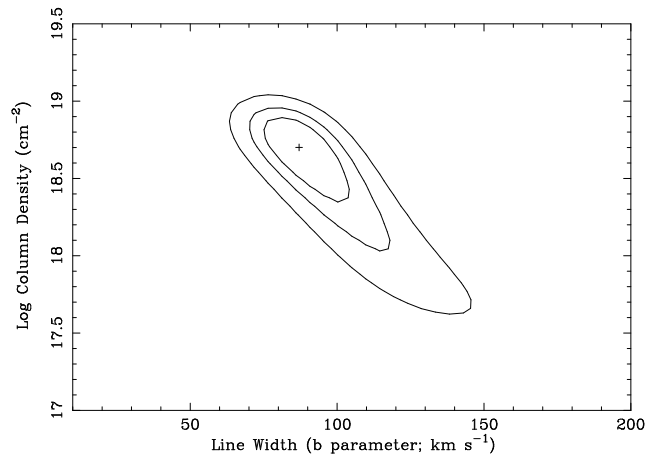}}}
\caption{Confidence region for column density and turbulent velocity
  for O\thinspace\textsc{vii} determined by fitting Voigt profiles to
  the O\thinspace\textsc{vii} $1s^2-1snp$, $n=2,3,4,5$ absorption
  lines.}
\label{ovii_voigt}
\end{figure}

O\thinspace\textsc{vii} is a prime candidate for analysis since four
lines in the first velocity zone are identified ($1s^2-1snp$,
$n=2,3,4,5$) in a region of the spectrum with high signal to noise.
The confidence region for turbulent velocity and column density
considering all four lines is shown in Fig.~\ref{ovii_1}. The column
density of O\thinspace\textsc{vii} in the first velocity zone is shown
to lie in the region $10^{18.36}$--$10^{18.64}$\pscm and the turbulent
velocity in the range 86--99\kmps with $1\sigma$. The implied optical
depth of the O\thinspace\textsc{vii} edge is 0.57 (taking the lower
1$\sigma$ bound of the column density) suggesting that
O\thinspace\textsc{vii} absorption is responsible for the majority of
the drop at 0.7~keV as predicted by the DWA model. This column density
places the $1s^2-1snp$, $n=2,3,4,5$ lines on the saturated part of the
curve of growth. The result was confirmed by fitting Voigt profiles
directly to the data (see section
\ref{confidence_region_determination} and Fig.~\ref{ovii_voigt}). As
mentioned by \citet{sako03} the presence of O\thinspace\textsc{vii}
forbidden emission suggests that the O\thinspace\textsc{vii}
$1s^2-1s2p$ may be partially filled in by resonance emission. To take
this into account the prescription of \citet{sako03} was used where
the ratio of forbidden to resonance emission is fixed at a value of 3
for a recombination-dominated plasma. One other issue is that the
$1s^2-1s5p$ transition lies over the position of the
Fe\thinspace\textsc{i} L$_3$ edge for Galactic absorption. By
calculating the equivalent width of the line-like feature at the edge
energy of Fe\thinspace\textsc{i} this can be taken into account. By
taking the limit of the confidence regions from all the various
combinations of corrections a more robust confidence region is
obtained. This produces a 1$\sigma$ confidence limit on the column
density of $10^{18.36}$--$10^{18.86}$\pscm and turbulent velocity of
79--100\kmps. If a similar analysis is performed for
O\thinspace\textsc{viii} then 1$\sigma$ confidence limits for column
density of $10^{17.42}$--$10^{18.47}$\pscm and for turbulent velocity
of 46--111\kmps are obtained.

\begin{figure}
\rotatebox{270}{
\resizebox{!}{\columnwidth}
{\includegraphics{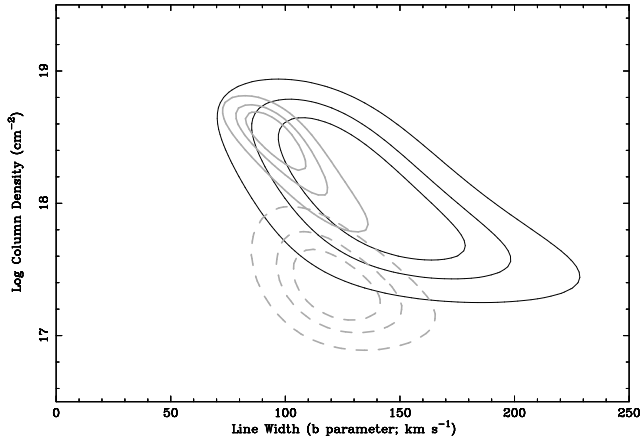}}}
\caption{Confidence regions for column density and turbulent velocity
  for O\thinspace\textsc{vii}. (\emph{black}): This paper's analysis
  using the 2000 data. (\emph{solid grey}): This paper's analysis
  using the 2001 data. (\emph{dashed grey}): \citet{sako03} analysis
  using the 2000 data.}
\label{comb_contour}
\end{figure}

\begin{figure}
\rotatebox{270}{
\resizebox{!}{\columnwidth}
{\includegraphics{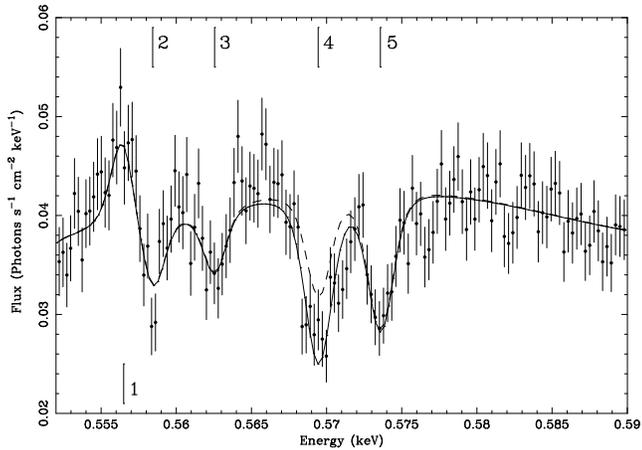}}}
\caption{Detail of spectrum around the O\thinspace\textsc{vii} $1s-2p$
  line. Overlaid is the best fitting model from Section
  \ref{line_parameter_determination} (\emph{solid}) as well as this
  best fitting model with the equivalent width of
  O\thinspace\textsc{vii} $1s-2p$ for the column density and turbulent
  velocity from \citet{sako03} (\emph{dashed}). Also labelled are
  (\emph{1}) the O\thinspace\textsc{vii} forbidden emission line,
  (\emph{2,3}) the O\thinspace\textsc{vi} KLL lines, (\emph{4}) the
  O\thinspace\textsc{vii} $1s-2p$ line for the first velocity zone and
  (\emph{5}) the O\thinspace\textsc{vii} $1s-2p$ line for the second
  velocity zone.}
\label{ovii_detail}
\end{figure}

The O\thinspace\textsc{vii} column density limit obtained above is
much larger than that obtained by \citet{sako03} using the earlier
\xmm\ observation. To investigate this issue the 2000 \xmm\ 
observation was analysed using the method described above. The best
fitting model from the 2001 data was used but the line and continuum
parameters were allowed to vary. Equivalent widths and errors for the
O\thinspace\textsc{vii} $1s^2-1snp$, $n=2,3,4,5$ lines were measured
as above and used to determine a confidence region for column density
and turbulent velocity for the 2000 data. This is shown in
Fig.~\ref{comb_contour} and is compared to the limits found by
\citet{sako03} and in this paper for the 2001 data. The contours
overlap the 2001 contours (which are much smaller because of the
higher signal-to-noise of the 2001 observation) but do not overlap the
contours found by \citet{sako03}. An examination of the equivalent
width obtained for the O\thinspace\textsc{vii} $1s^2-1s2p$ line
reveals a significant difference between that found here (1.14~eV,
44~m\angstrom) and that found by \citet{sako03} (0.48~eV,
18~m\angstrom). We are unable to account for this difference. The
region around the O\thinspace\textsc{vii} $1s^2-1s2p$ line is shown in
Fig.~\ref{ovii_detail}. Also shown is the best fitting model from
Section \ref{line_parameter_determination}, as well as this model with
the O\thinspace\textsc{vii} $1s^2-1s2p$ line with the equivalent width
expected using the column density and turbulent velocity of
O\thinspace\textsc{vii} determined in \citet{sako03}.  This equivalent
width also includes the expected response emission infilling.  Clearly
the \citet{sako03} values of column density and turbulent velocity
under-predict the equivalent width of the line.

\subsubsection{Carbon}

C\thinspace\textsc{vi} has five identified lines ($1s-np,
n=2,3,4,5,6$). Calculating the confidence region for this species
produces a poor fit and two separate $\chi^2$ minima because the $n=3$
line has a lower equivalent width than the $n=4$ line when it should
have a larger equivalent width due to its higher oscillator strength.
There are two possibilities to explain this: either the $n=3$ is
underpredicted (because unidentified low equivalent width lines nearby
mean the fitted continuum is lower than the true continuum) or the
$n=4$ transition is blended with another transition (so increasing its
apparent equivalent width).  Since no other transition can be found
that fits the position of the $n=4$ transition the former explanation
is favoured. Removing the $n=3$ line from the analysis results in a
single minimum in the confidence region with column density lying in
the range $10^{16.46}$--$10^{16.77}$\pscm and turbulent velocity in
the range 122--153\kmps (1$\sigma$ confidence limit). Because no
measurement is available for the strongest oscillator strength
transition ($1s^2-1s2p$) in C\thinspace\textsc{v}, the column density
and turbulent velocity could not be tightly constrained. 1$\sigma$
confidence limits were obtained of $10^{15.91}$--$10^{19.47}$\pscm for
column density and 48--130\kmps for turbulent velocity.

\subsubsection{Nitrogen}

The absorption lines of N\thinspace\textsc{vii} prove more difficult
to understand. Only one line is detected (Ly$\alpha$) but by using an
upper limit on the equivalent width of the undetected Ly$\beta$ line a
confidence region for column density and turbulent velocity can be
determined. This region, however, does not agree with the region
obtained by directly fitting a Voigt profile to the Ly$\alpha$ line.
A possible explanation for this is that the Ly$\alpha$ line is not a
single line but made up of several lines with a velocity separation
greater than the turbulent width of the lines, but less than the
instrumental resolution. Looking at the abundance of individual
species in each zone of the DWA models presented in \citet{turner03}
it is apparent that while other species have their column densities
preferentially in one zone, N\thinspace\textsc{vii} has its column
density more evenly distributed between the zones. Considering the
N\thinspace\textsc{vii} Ly$\alpha$ line to be made up of multiple
lines from different zones brings the limit on column density into
reasonable agreement with the predictions of the DWA models.

\subsubsection{General}

\begin{figure}
\rotatebox{270}{
\resizebox{!}{\columnwidth}
{\includegraphics{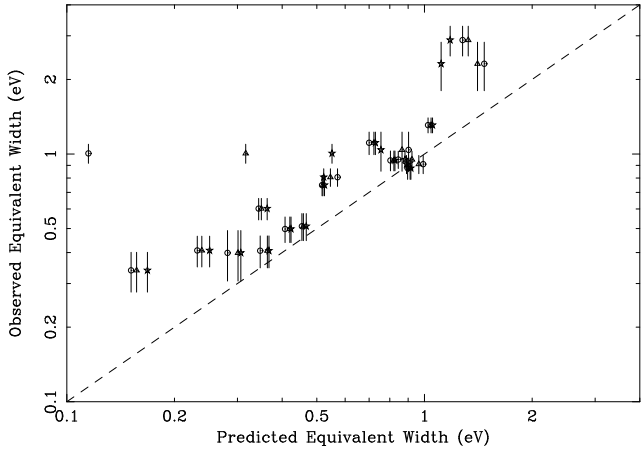}}}
\caption{Comparison of equivalent widths for absorption lines between
  those observed and those predicted by the \cloudy\ DWA models from
  \citet{turner03}. Circles indicate model 1, triangles model 2 and
  stars model 3 from \citet{turner03}. Only identified lines with
  accurate line details from \citet{verner96} are included.}
\label{cloudy_comp}
\end{figure}

Table \ref{tab_comp} compares the column densities and turbulent
velocities derived above with those of the DWA models from
\citet{turner03}. As can be seen reasonable agreement is found between
the values empirically derived here and those obtained by fitting DWA
models to the data. Finally we compare the observed equivalent widths
to the equivalent widths predicted by the DWA models. As can be seen
from Fig.~\ref{cloudy_comp} a reasonable match is found although the
observed absorption lines are found to be larger than predicted by a
factor of $\sim1.25$. This may be due to a combination of different
species having different abundances than the assumed Solar abundance
or the gas having a different turbulent velocity than the assumed
100\kmps. For example O\thinspace\textsc{vii} requires an increase in
abundance of $\sim3-5$ or an increase in turbulent velocity from 100
to $\sim$130~\kmps to account for the $\sim1.25$ difference between
observed and predicted equivalent widths.

\begin{table}
\centering
\caption{Comparison of column densities ($N_\mathrm{X}$) and turbulent
velocities ($V_\mathrm{turb}$) between this work and DWA models in
\citet{turner03}. $N_\mathrm{X}$ is given in units of $\log
\mathrm{cm}^{-2}$ and $V_\mathrm{turb}$ in units of \kmps. The DWA models in
\citet{turner03} assumed a $V_\mathrm{turb}$ of 100 \kmps.}
\begin{tabular}{lccccc}
\hline
Species  & \multicolumn{3}{c}{DWA models} & \multicolumn{2}{c}{This
  work} \\
 & 1 & 2 & 3 & & \\
\hline
 & $\log N_\mathrm{X}$ & $\log N_\mathrm{X}$ & $\log N_\mathrm{X}$ &
 $\log N_\mathrm{X}$ & $V_\mathrm{turb}$ \\
 & \pscm & \pscm & \pscm & \pscm & \kmps \\
\hline
\hline
O\thinspace\textsc{vii} & 18.12 & 18.18 & 18.19 & 18.36--18.86 & 79--100  \\
O\thinspace\textsc{viii} & 18.12 & 18.19 & 17.93 & 17.42--18.47 & 46--111  \\
C\thinspace\textsc{v} & 17.74 & 17.82 & 17.80 & 17.27--19.65 & 122--153 \\
C\thinspace\textsc{Vi} & 17.52 & 17.57 & 17.57 & 17.63--17.84 & 48--130  \\
\hline
\end{tabular}
\label{tab_comp}
\end{table}

\section{Absorption by dust}

\label{dust}

The presence of dust in \mcg\ was first suggested by
\citet{reynolds97} to explain the large Balmer decrement found in the
optical spectrum. \citet{lee01} found that the drop in flux at
$\sim0.7$~keV had the exact shape of absorption by
Fe\thinspace\textsc{i} and that the $N_H$ implied by the discontinuity
at the Fe\thinspace\textsc{i} L-edge was consistent with that
determined from $E(V-B)$ reddening. This shape has also been seen in
the Galactic black hole candidate Cygnus X-1 \citep{schulz02}.
\citet{lee01} attributed this absorption to dust consisting of iron
oxides or silicates (such as olivine) embedded in the warm absorber.
However the sensitivity of the HETGS at the neutral oxygen edge was
insufficient to measure the column of neutral oxygen accurately. Using
RGS data \citet{sako03} produced an upper limit at 90 per cent
confidence on O\thinspace\textsc{i} column density of $10^{16}$\pscm
using the apparent absence of a drop across the O\thinspace\textsc{i}
K-edge at 0.538~keV. Such a column density corresponds to a 0.5 per
cent drop in flux across the O\thinspace\textsc{i} edge.

Using our higher quality RGS data we have measured the drop in flux
across the neutral O\thinspace\textsc{i} edge. The 0.5--0.555~keV
region was fitted to a power-law, Galactic absorption, the line
profiles found in section \ref{warm_params} and an absorption edge at
the O\thinspace\textsc{i} edge energy. This produces an
O\thinspace\textsc{i} edge optical depth of 0.16--0.23 ($1\sigma$)
corresponding to a column density of $10^{17.51}-10^{17.67}$\pscm.
Given the column density of Fe\thinspace\textsc{i} of $10^{17.3}$\pscm
found in \citet{turner03}, this suggests a neutral Fe:O ratio of
$\sim$1:2.  This is consistent with the dust being in the form of iron
oxides or silicates as found by \citet{lee01}.

We repeated the above analysis on the 2000 data. Using 90 per cent
confidence limits we find an optical depth at the edge of 0.13--0.34,
consistent with the result found with the longer observation. As well
as finding a different optical depth across the edge, the 90 per cent
confidence limit range found here is an order of magnitude larger than
that found by \citet{sako03} using the same data.

\section{Line variability}

\label{variability}

\subsection{Short timescale variations}

\label{short}

To test for short term variability in the warm absorber the
observation was split into two sets; one when the flux level was below
a certain level and one when it was above that level. The level was
set so an equal number of counts was present in each set. The level
was found to be 16.6\ctsps for the 0.2--2~keV EPIC pn light-curve. The
data from each set were then reduced using the method described in
section \ref{data_reduction}. The best fitting continuum and line
model from section \ref{warm_params} was fitted to each set of spectra
separately and the errors for the best fitting parameters calculated
as before. No significant change in position of the lines was found.
Only 4 lines have a ratio of the equivalent widths from the high flux
period to the low flux period which individually differs significantly
from 1 (greater than $2\sigma$; lines 20, 21, 36 and 46 corresponding
to O\thinspace\textsc{i}, O\thinspace\textsc{ii},
Fe\thinspace\textsc{xvii} and Fe\thinspace\textsc{xxi} with
significances of 3.0$\sigma$, 3.0$\sigma$, 4.8$\sigma$ and 2.0$\sigma$
respectively). Given the large number of lines found, it is likely
that random fluctuations would give false variability detections for
one or two lines at a 2$\sigma$ threshold (the false alarm probability
for 51 lines at 2$\sigma$ individually is $\sim$0.4). Since three of
the lines have higher detection significances than this, the
variability in these lines is likely to be real. Individually all
other lines have a ratio that is consistent with 1. The variability in
line strength can be used to estimate an upper limit for the distance
between the warm absorbing region and the central source.  Assuming
that the increase in equivalent width is produced by recombination of
more highly ionized species (brought about by the drop in continuum
flux) then the recombination timescale,
$\tau_{\mathrm{rec}}\lesssim\tau_{obs}$, the timescale over which the
spectrum is observed to change. $\tau_{obs}$ is taken to be the
average length of bright and dim time intervals, which is
$\sim2500$~s. From \citet{shull82}
\begin{equation}
\tau_{\mathrm{rec}}=\frac{A_{rad}T^{\chi_{rad}}}{n}
\end{equation}
where $T$ is the temperature of the gas, $n$ is the density and
$A_{rad}$ and $\chi_{rad}$ species dependant parameters. Using the
definition of the ionization parameter used in \citet{turner03}
\begin{equation}
\xi=\frac{L_{2-10}}{n R^2}
\end{equation}
 then
\begin{equation}
\tau_{\mathrm{rec}}=\frac{A_{rad}\xi R^2T^{\chi_{rad}}}{L_{2-10}}.
\end{equation}
Consequently, if $\tau_{rec}\lesssim\tau_{obs}$ then
\begin{equation}
R\lesssim \sqrt{\frac{\tau_{obs} L_{2-10} A_{rad}}{\xi T^{\chi_{rad}}}}.
\end{equation}
The O\thinspace\textsc{i} and O\thinspace\textsc{ii} lines are
produced solely in the low ionization zone. Taking the appropriate
values for $\xi$, $L_{2-10}$, $T$ from \citet{turner03} gives an upper
limit of the radius of this zone as $10^{19.5}$\cm.
Fe\thinspace\textsc{xvii} and Fe\thinspace\textsc{xxi} are found
predominately in the highly ionized zones and using the appropriate
values for this zone constrains it to lie within $10^{17.0}$\cm\ of the
central source.

\begin{figure}
\rotatebox{270}{
\resizebox{!}{\columnwidth}
{\includegraphics{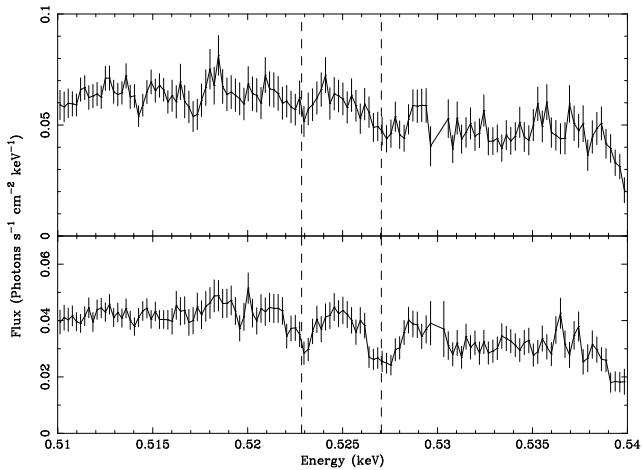}}}
\caption{Detail of RGS spectra for high flux period (\emph{upper}) and
  low flux period (\emph{lower}). The dashed lines show the position
  of the O\thinspace\textsc{i} and O\thinspace\textsc{ii} absorption
  lines. Their equivalent width is clearly higher during the low flux
  period.}
\label{oxvari}
\end{figure}

\subsection{Long timescale variations}

To test the longer timescale variability in the warm absorber, the
data from a previous observation of \mcg\ with \xmm\ was extracted
using the same process as described in Section \ref{data_reduction}.
The line and continuum model of Section \ref{warm_params} was fitted
to the data and the resulting parameters compared to the values
obtained from the 2001 data analysed here. Because of the lower
signal-to-noise ratio of the 2000 data not all the lines are found, so
a comparison is only possible on the strongest lines. No lines appear
to have statistically significant changes, possibly due to the poorer
signal to noise in this data set.  Examination of the spectra,
however, reveals that several lines appear in the \citet{brand-ray01}
spectrum but have fallen below a detectable level a year later. These
lines occur at $0.510715\pm0.0003$ and $1.12631\pm0.001$~keV and are
identified as Ca\thinspace\textsc{xiv} and Fe\thinspace\textsc{xxiii}
respectively.

\section{Discussion}

\label{discussion}

The analysis of the soft X-ray absorption lines of \mcg\ have revealed
some interesting properties of the warm absorber. Most importantly the
column densities measured for O\thinspace\textsc{vii} and
O\thinspace\textsc{viii} lines are large enough, when considered with
Fe\thinspace\textsc{i}, to account for the drop in flux at
$\sim$0.7~keV. This means that the dominant effect on the shape of the
soft X-ray spectrum of \mcg\ is from a dusty warm absorber. Any
additional complexity from relativistically blurred emission lines is
of low equivalent width.

The column densities and turbulent velocities that were measured for
the other species agree with the predictions of the DWA model
presented in \citet{turner03}. The measured equivalent widths of the
identified lines are larger than the predicted ones from the DWA
models of \citet{turner03} but only by a factor of $\sim$1.25 which
may be accounted for by a slightly higher than Solar abundance or
higher than 100\kmps turbulent velocity in the warm absorbing gas than
that assumed in the DWA models.

A broad range of ionization states is found for
several elements, most notably oxygen and iron. This suggests the warm
absorber consists of many zones with a range of ionization parameters
as was assumed in \citet{turner03} or possibly a continuum of
ionization parameter as suggested by \citet{morales00}.

The variability of various absorption lines suggests that the most
lowly and highly ionized gas is varying as the continuum flux varies.
The increase in equivalent widths of these lines with decreasing flux
is presumably from recombination of higher species in response to the
drop in ionising flux. Using values for the recombination timescale
upper limits can be placed on the distance of these two zones from the
source of ionizing flux. The low ionization zone gas is found to line
within $\sim$10 pc of the central source whereas the highly ionized
zone must lie very close to the central engine at $10^{17}$\cm, at a
similar distance as the BLR.

\section*{Acknowledgments}

Based on observations obtained with \xmm, an ESA science mission with
instruments and contributions directly funded by ESA Member States and
the USA (NASA). AKT acknowledges support from PPARC. ACF thanks the
Royal Society for support. JCL thanks the Chandra fellowship for
support. This was provided by NASA through the Chandra Postdoctoral
Fellowship Award number PF2-30023 issued by the Chandra X-ray
Observatory Center, which is operated by SAO for and on behalf of NASA
under contract NAS8-39073.

\bibliographystyle{mnras}                       %% MNRAS
\bibliography{mn-jour,akturner_may25}

\end{document}